# Metasurfaces for suppressing reflection over broadband

Anuradha Patra, Ajith P Ravishankar, Arvind Nagarajan and Venu Gopal Achanta

*Tata Institute of Fundamental Research, Homi Bhabha Road, Mumbai 400005 INDIA*

Surfaces patterned with arrays of quasi-periodic air holes having conical depth profile have been studied for their effectiveness in suppressing air-substrate reflection in the wavelength range of 450-1350 nm (limited by our measurement). The role of quasi-periodic air-hole pattern, depth of holes and launch angle on the observed antireflection behavior are investigated. The average optical transmittance of the patterned quartz substrate at near normal incidence is more than 97% and reflectance is less than 2%. Patterned quartz surfaces with 450 nm thin graded rarefaction region maintain the antireflective property up to 30° (limited by our measurements) angle of incidence.

Anireflection coatings are well studied starting with the work of Lord Rayleigh on graded index medium[1]. While single layer coatings with refractive index $n_1 = \sqrt{n_0 n_2}$ is used to suppress the reflection between layers with refractive index $n_0$ and $n_2$, quarter wave stacks, on the other hand, effectively cancel the reflections from each layer by destructive interference[2,3]. Although graded index medium is known to work over broadband[4,5], the main problem is in realizing an optical material that has refractive index close to that of air (*n*=1). Porous materials used in graded index layer based antireflection coatings are shown to have refractive index[5] as low as 1.05. An alternative to these techniques is to pattern the surface with dimensions smaller than the wavelength of light for low reflection. These metasurfaces, nano-patterned surfaces with sub-wavelength tapered features, provide adiabatic refractive index matching between successive layers. An effective media with an axially varying effective refractive index may provide a graded transition of refractive index between the air and substrate, and consequently eliminate the undesirable reflection by blurring an abrupt interface[6-8]. The best known natural examples are the corneas of moth and butterfly eyes and the transparent wings of hawk moths[9]. In antireflective surfaces mimicking the nature, tapered nanorods are typically used on different substrates such as quartz, Si and polymer[10-12]. In nano-patterned substrates approach, where the refractive index changes gradually from air- to substrate-like, Liu et al[13] have demonstrated reduced reflection of 5% by nano-patterning Si substrate to different thicknesses. A new class of materials known as "reflectionless potentials" having a predefined refractive index profile is demonstrated recently to have truly broadband and omni-directional antireflection property[14]. Similar response is also demonstrated in the so-called "all-dielectric metamaterials"[15].

Unlike depositing layers on top of the substrate, one may also pattern the substrate to achieve the graded index profiles. For example, the role of geometrical effects in antireflection property of sub-wavelength gratings was studied by Song and Lee[16]. In this context, compared to periodic patterns, quasi-periodic patterns offer several advantages like broadband response that is polarization and launch angle independence. Recently, broadband enhanced transmission that is launch angle and polarization independent as well as possessing designable spectral response are reported in plasmonic quasicrystals[17]. In this paper we report suppression of reflection of unpolarized light to as low as 2% over 450-1300 nm wavelength range that is launch angle independent. Designable spectral response as well as launch angle independence makes these suitable for light harvesting and other photonic applications.

Quartz substrates of 500μm thickness have been patterned to have quasi-periodic air hole arrays by electron beam lithography followed by a combination of dry and wet etching processes to create sub-wavelength deep tapered holes to gradually vary the refractive index from air-like at the top to substrate-like at the bottom. The quasi-periodic pattern is designed by a modified oblique tiling method where the set of parallel lines are replaced by 2D arrays of dots. The separation between the dots is the base period that defines the spectral response. The initial 2D array of dots is rotated by $\frac{\pi}{5}$ angle each time and superposed 4 times with a common center. The intersection points of the 5 sets of these 2D arrays of dots correspond to the quasi-crystal lattice with 5-fold rotation symmetry. Details and the advantages of this method are reported in a recent work[17]. In order to realize the tapered holes on insulating substrates, to avoid charging effects during electron beam lithography, a 100 nm thick sputtered Aluminium (Al) mask was used. This was followed by spin coating PMMA 495 A4, an electron beam resist, onto Al film. The thickness of the resist was ~200 nm. The resist was then patterned to have quasi-periodic structure by electron beam lithography using Raith e-line. The sample was developed in a 1:3 solution of methyl isobutyl ketone (MIBK) and isopropyl alcohol (IPA). The pattern was then transferred to Al mask by dissolving the exposed Al in MF26A (a photoresist developer). Reactive ion etching (RIE) was then employed to strip off the resist by using oxygen plasma with a RF-power of 80 W. Thereafter, etching of quartz substrate was done in $CHF_3:O_2$ (98:2) ambient at RF-power of 150 W. Finally, the Al mask was removed by rinsing in MF26A and DI water.

The samples studied are with base periods of 600 nm (Sample-1) and 1000 nm (Sample-2) and a third one with a bigrating pattern with two different inherent periods of 600 nm and 1000 nm combined with a common center (Sample-3). In order to investigate the dependence of the reflectance on the effective layer thickness, three samples of each base period are prepared by using different etching times. RIE for longer durations generally gives tapered sidewalls[18,19]. The RIE-induced tapered hole depths, measured by atomic force microscopy, were found to be (a) 200±07 nm (b) 325±05 nm and (c) 450±10 nm. Angle resolved white light transmission and reflection measurements are performed with a near collimated light from a 100 W tungsten halogen lamp with <3° divergence as a source and fiber coupled spectrometers (Ocean Optics S4000 and NIRQUEST).

Fig. 1, Fig. 2 and Fig. 3 summarize the results for the sample-1, sample-2 and sample-3, respectively. The SEM image of quasi-pattern (on a metal substrate) formed with initial rhombus having 1 mm long sides having holes at a periodic separation of 600 nm is shown in Fig. 1a. It may be seen that the quasicrystal pattern has no short range periodicity as expected. The average hole diameter in the structure is about 90±08 nm. Fig. 2a is the SEM image of the quasicrystal structure with 1000 nm separation between the holes in the base array. Fig. 3a is the image of the bigrating quasi-crystal structure obtained by superposing the 600 nm and 1000 nm patterns with a common center. The diffraction pattern of these quasi-periodic structures exhibits $\frac{\pi}{5}$ rotational symmetry and dense k-space[17].

Figs. 1 b-d are, respectively, the plots of the reflectance spectra of sample-1 with different air hole depths. We can infer from Figs. 1 b-d that at near-normal incidence the reflectance for all the samples is < 2% over the 450-1350 nm wavelength range. However, the difference in reflectance spectra arises at higher angles of incidence. As the angle of incidence increases, a gradual increase in reflection is noticed for the shallow pattern (Fig. 1b) and a less steeper increase for 325 nm deep structure (Fig. 1c). However, the quasi-periodic hole patterned quartz substrate with ~450 nm deep holes show a weaker variation in reflectance over the wavelength range of 450-1350 nm for 0° to 30° angle of incidence range (Fig. 1d). The results presented are limited by our measurement geometry and better performance is expected in both wavelength and launch angle ranges. Similar observations are seen in sample-2 (Figs. 2 b-d) and sample-3 (Figs. 3 b-d) as well but with reduced dependence on the incidence angle in both the samples. Sample-3, as seen in Fig. 3c, shows a slight variation in reflectance when the patterned surface is only ~325 nm deep. In gradient index optics, the magnitude of reflection depends on the depth of the graded index as well as on the smoothness of the optical path inside the film[3,4,20,21]. The gradual the curvature of light within the layer i.e. smoother the transition of the refractive index from free space ($n$=1, air-like) to a value that matches the index of the substrate (substrate-like), lower is the reflection over the whole spectrum. The gradual curvature of light (refractive angle) is highly susceptible at higher angles of incidence where a tiny increment of refractive angle is angle is perceived as a large variation of the characteristic index[4,20,21]. As a result, a considerable deformation of the index profile is seen by the incoming light. This leads to higher reflectance at off normal incidence angles. To overcome this, graded index coatings with specific index index profiles such as quintic, gaussian, exponential, sinusoidal, exponential-sine, klopfenstein etc. have been studied in literature[22-26]. In general, the effective index changes more slowly when the depth is increased which furthermore leads to decrease in reflectance. Hence, an optimum depth so as to ensure a least variation of refractive angle is necessary to avoid abrupt interface. For sample-1 (Fig. 1d) and sample-2 (Fig. 2d), a minimum of ~450 nm gradient effective medium is necessary to achieve low reflectance at off-normal incidence. However, for sample-3 an effective thickness of ~325 nm is sufficient to provide a broadband and angle independent low reflectance (Fig. 3c).

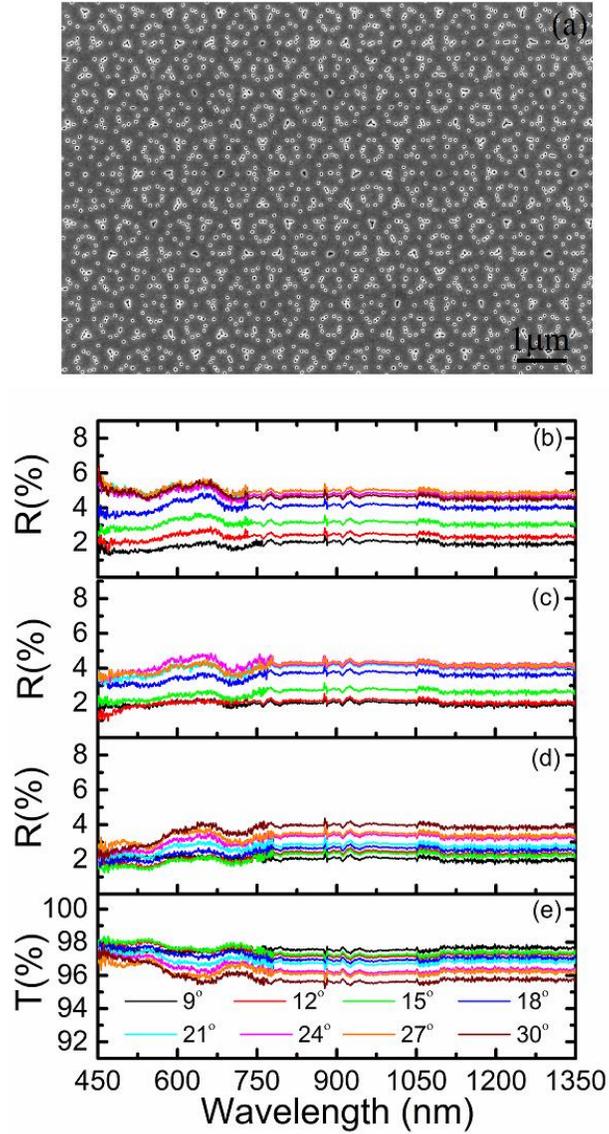

FIG.1. (a)SEM images of sample-1. Reflectance spectra of sample-1 with the effective thickness of the layer of graded refractive index (b) ~200 nm (c) ~325 nm (d) ~450 nm (e) Transmittance spectra of sample-1 with ~450 nm thick graded refractive index.

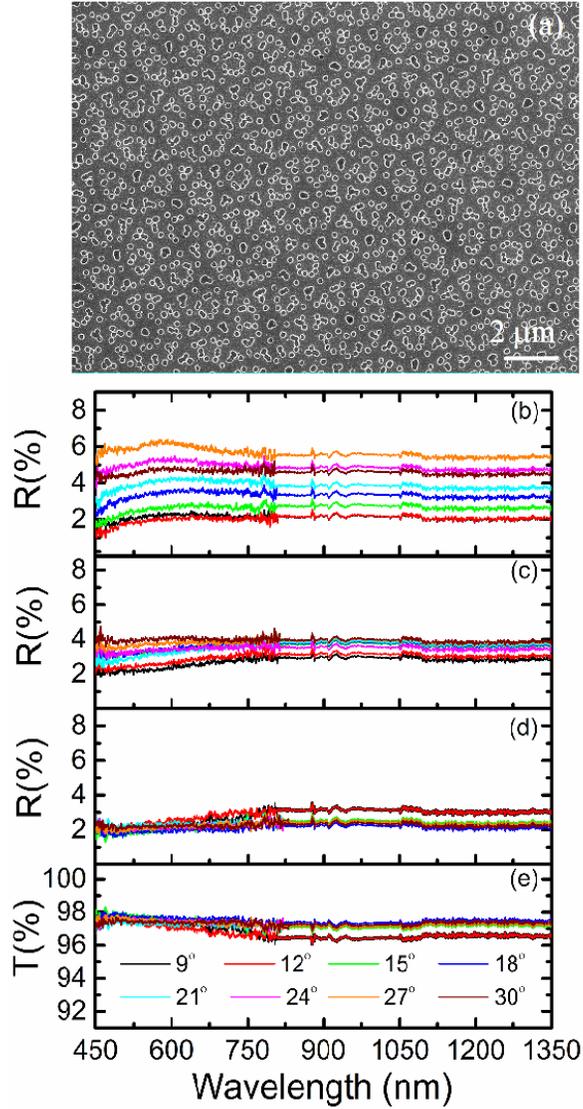

FIG.2. (a)SEM images of sample-2. Reflectance spectra of sample-1 with the effective thickness of the layer of graded refractive index (b) ~200 nm (c) ~325 nm (d) ~450 nm (e) Transmittance spectra of sample-2 with ~450 nm thick graded refractive index.

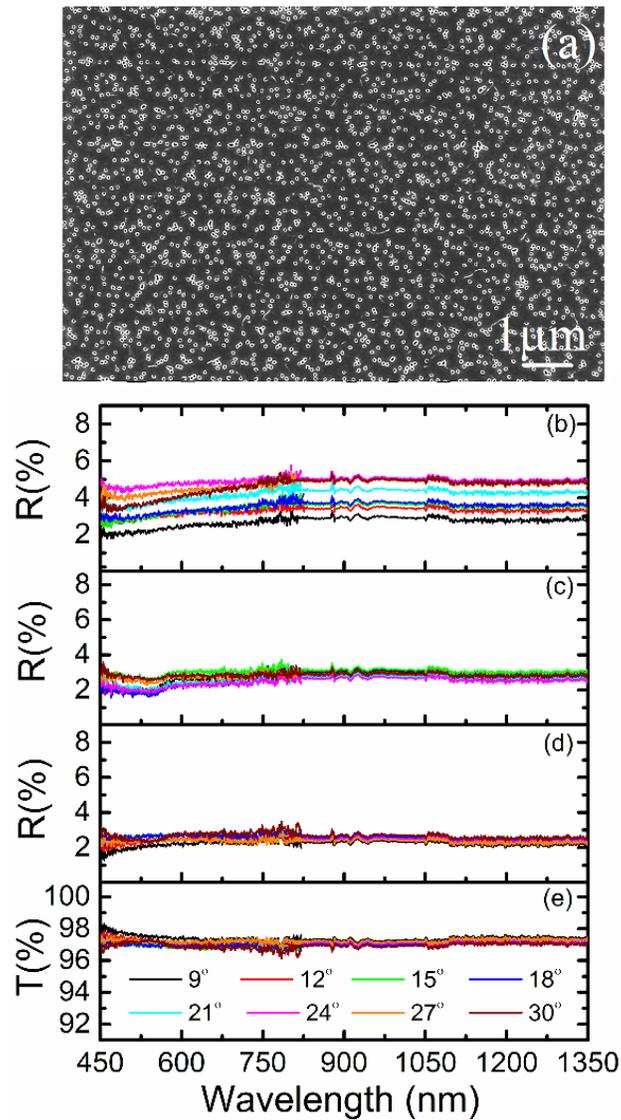

FIG.3. (a)SEM images of sample-3. Reflectance spectra of sample-3 with the effective thickness of the layer of graded refractive index (b) ~200 nm (c) ~325 nm (d) ~450 nm (e) Transmittance spectra of sample-3 with ~450 nm thick graded refractive index.

In order to understand the basic reason behind different antireflective behaviors of the three samples, it is important to note that the height and shape of graded index layer as well as period are some of the key factors that govern the light trajectory in metasurface. A tapered profile, in principal, can be regarded as an infinite stack of infinitesimally thin layers[4]. Owing to gradual tapering, each infinitesimally thin layer has gradual difference in volume fraction (of air) and therefore a gradual change in refractive index between successive infinitesimally thin layers is obtained. Though the air-holes in all the samples are of ~90 nm in diameter, the three samples have different density of air holes and thus fill factors. For instance, on top surface, Sample-1,-2,-3 have $n_{eff}$ as 1.38, 1.42 and 1.35, respectively. Subsequently, the refractive index profiles of each of the samples are different and so is their respective antireflective property.

In addition to the effective refractive index layers, the role of quasi-periodic arrangement of air holes is also very important. For example, considering that the fill factor of sample-1 is more than that of sample-2, one would expect the response of sample-1 to be better than that of sample-2. However, the antireflective performance of 1 μm base period quasi-periodic structure is better than that of quasi-periodic structure with base period 600 nm. Hence, an appropriate choice of the base period is necessary. Transmittance spectra of sample-1, sample-2 and sample-3, with ~450 nm deep effective graded index are plotted in Figs. 1e, 2e and 3e, respectively. It is seen that the samples show high transmission and therefore it can be concluded that there is no appreciable scattering loss from these surfaces. It is necessary to note that unlike other periodic arrangements, quasi-periodic arrangement can possess higher density of air-holes and therefore has denser *k*-space. In fact, owing to the dense *k*-space, the quasi-periodic patterned quartz substrate shows broadband, omnidirectional antireflective characteristic with enhanced light transmission.

In Fig. 4, theoretical reflectance plot of a 100-layer graded index medium with refractive index of each layer (*n*) being equivalent to that of Sample-3c (*n* varies from 1.35 to 1.45) is shown. It is seen that the graded index thin film stack displays similar antireflection characteristics but have much larger angle dependence compared to quasi-periodic air hole patterned quartz. There is almost 2% change in reflectance when the angle of incidence is varied from 0-40°. This is a manifestation of the role of air hole quasi-periodic arrangement in the observed antireflection property. Substrates patterned with tapered air-holes arrays conforming to quasi-periodic lattice show potential in suppressing Fresnel reflection over a wide range of incidence angles and wavelengths.

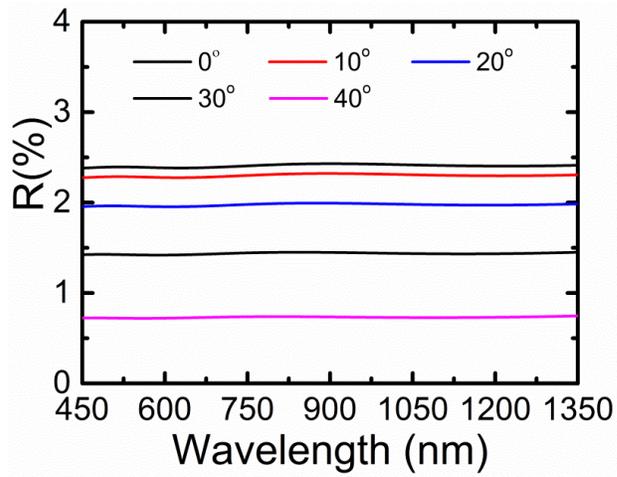

FIG. 4. Theoretically calculated reflectance plot of 100-layer graded thin film with refractive index gradually varying from 1.35 to 1.45.

In conclusion, metasurfaces realized by patterning quartz substrates with tapered air hole arrays conforming to 5-fold rotation symmetry quasi-periodic lattice are demonstrated to have broadband and omni-directional anti-reflective properties. High transmission (>97%) and reduced reflection (<2%) in the wavelength range of 450-1350 nm and launch angle range of 0-30° (limited by our measurements) are observed. Three quasi-periodic structures with three different air hole depths were investigated. The air hole arrangement and depth of the tapered holes play a crucial role in determining the reflection characteristics, in particular, at oblique angles of incidence.